\documentclass[prl]{revtex4}%
\pdfoutput=1

\usepackage{epsfig}
\usepackage{graphics}

\begin{document}

\title{Holographic Indeterminacy, Uncertainty  and  Noise}

\author{Craig J. Hogan}
\affiliation{University of Washington,  Seattle, WA 98195-1580, USA}

\begin{abstract} 
A  theory is developed to describe the nonlocal effect of spacetime quantization on position measurements transverse to macroscopic separations.  Spacetime quantum states close to a classical null trajectory are approximated by plane wavefunctions of Planck wavelength ($l_P$) reference beams; these are used to connect transverse position operators at macroscopically separated events.     Transverse positions of events with null spacetime separation, but separated by macroscopic   spatial distance $L$,  are shown to be quantum conjugate observables, leading to  holographic indeterminacy and a new uncertainty principle, a lower bound on the standard deviation of  relative transverse position $\Delta x_\perp > \sqrt{l_PL}$ or angular orientation 
$\Delta\theta > \sqrt{l_P/L}$.  The resulting limit on the number of independent degrees of freedom  is shown to   agree quantitatively with  holographic covariant entropy bounds   derived from black hole physics and string theory.   
The theory  predicts   a   universal  ``holographic noise'' of spacetime, appearing as shear perturbations with a frequency-independent power spectral density $S_H=l_P/c$, or in equivalent metric perturbation units, $ h_{H,rms}\simeq \sqrt{l_P/c} = 2.3 \times 10^{-22} /\sqrt{\rm Hz}$.  If this description of holographic phenomenology is valid, interferometers with current technology could undertake direct quantitative studies of quantum gravity.
\end{abstract}
\maketitle
\section{introduction}
In standard field theory, quantum particles and fields propagate and interact on an unquantized, classical spacetime manifold.   Their behavior  is described in terms of  quantum operators that are functions of spacetime position. 
This idealized model  entirely neglects the quantum degrees of freedom of spacetime itself. 
In reality, spacetime  is widely thought\cite{Seiberg:2006wf} to be   a quantum system   whose apparent classical properties, including  the fundamental and invariant causal structure defined by null paths (such as  light rays), emerge as a limit of a more fundamental quantum system.

Longstanding approaches to quantum gravity (as in ref. \cite{gibbons}) based, like field theory,  on  local quantization analysis, suggest  that effects of spacetime quantization only become important at the Planck scale, $l_P=\sqrt{\hbar G_N/c^3}= 1.616\times 10^{-33} {\rm cm} $,  and average out on larger scales or lower energies. However, a reasonable alternative hypothesis is that positions on the spacetime manifold, like any other classical quantities,  are defined by  quantum observable operators. 
In that case,   emergent spacetime will ultimately be described by introducing a fundamental  quantum system  prior to the assignment of spacetime position observables.     In   an extended, emergent  ``spacetime made of waves'',  nonlocal measurements, involving comparison of observables at widely separated spacetime events, can show  new  behavior due to Planck scale quantization   that is not predicted in field theory.  

   Indeed, general arguments\cite{Hogan:2007rz,Hogan:2007hc}, based on the idea of the spacetime metric emerging from a   wavefunction at Planck resolution, suggest  a  new quantum behavior of spacetime, called holographic indeterminacy, that may expose its quantum degrees of freedom to direct experimental investigation. The wavefunction of a spacetime null path  connecting two events is modeled as a plane wave of Planck wavelength, with uncertain orientation for particles localized in the transverse directions.
    Positions transverse to null trajectories are  defined by quantum observables relative to hypothetical reference beams of Planck-wavelength radiation that define the metric.  This general and simple idea is only  an effective theory of a new quantum behavior,  rather than  a fundamental theory,  but is sufficient to estimate the nonlocal, macroscopic physical phenomenon of holographic indeterminacy:   
nonlocal measurements of  transverse  positions display nonclassical quantum indeterminacy on scales much larger than the Planck length.     This paper develops more fully the theory of   holographic indeterminacy of spacetime and its observable consequences.

Independent motivation for holographic indeterminacy comes from  the fact that the  resulting decomposition of spacetime eigenstates, while not in agreement with field theory, does  agree with holographic behavior  of spacetime degrees of freedom estimated from the physics of black hole evaporation  and string theory\cite{Bekenstein:1972tm,Bardeen:gs, Bekenstein:1973ur,Bekenstein:1974ax,Hawking:1975sw,'tHooft:1985re,Strominger:1996sh,'tHooft:1993gx,Susskind:1994vu,'tHooft:1999bw,Bigatti:1999dp,Bousso:2002ju,Maldacena:1997re,Witten:1998zw,Aharony:1999ti,Alishahiha:2005dj,Horowitz:2006ct,Padmanabhan:2006fn}. Holographic properties, including the scaling of entropy with area instead of volume and  nonlocality of quantum correlations, have long been in puzzling contradiction to field theory.  Holographic indeterminacy accounts for these effects, and also defines the spacetime states more explicitly than string theory.  Whereas string theory  has not had a clearly developed classical limit to show the character of the  effects of holography  on observable phenomena viewed from inside a nearly-flat spacetime, 
 the concrete hypothesis of holographic indeterminacy--- which can be regarded as a particular hypothesis for how holographic encoding works in nature--- results in definite predictions of observable physical effects.  The theory presented here   describes  one way  holographic behavior could appear to observers in a nearly flat spacetime like our local environment.

Holographic indeterminacy is  thus distinguished  from previous treatments of holographic effects by being a more specific hypothesis with a more concrete predicted phenomenology. In other respects  holographic indeterminacy appears to be a rather conservative implementation of holography. Quantum mechanical unitarity is preserved, as are local symmetries; the only new effects are nonlocal.  Moreover, the nonlocal effects are localized in angle; the new correlations and holographic noise grow  gradually from events along null trajectories and stay close to the classical trajectories. No new particles, symmetries, or dimensions are predicted, and indeed no new connection is made to physical fields; the new behavior is entirely geometrical and entirely nonlocal.  Even Planck's constant cancels out of the final  results aside from   one fundamental length, the Planck length.     It remains to be seen how to apply these ideas   in highly curved spacetimes such as the regions near black hole or cosmological event horizons; the analysis here deals only with nearly flat spacetimes.

It is interesting that the most promising experimental probes of this Planck-scale physics do not use Planck energy particles directly.  Instead, the probes  are macroscopic elements of laboratory-scale interferometers--- proof masses whose own wavefunctions are spatially narrow and remain so even under a precise position measurement, so that the uncertainty principle allows  measurement of the small spread in spacetime wavefunction.  It is an open question   whether precision studies  of atomic-scale quantum systems  can also be used to probe the effect.

\section{holographic indeterminacy in emergent spacetime}

Holographic indeterminacy results from diffraction of the fundamental Planck scale waves over macroscopic distances. A  classical trajectory is specified by  the orientation of a plane wave (propagating on the $z$ axis, say) and a transverse position $(x_\perp,y_\perp)$ in the plane of the wave.  However, specifying a transverse position creates an uncertainty in transverse momentum, and therefore in distant transverse position on the same axis. Thus position observables at widely separated points in an emergent spacetime become conjugate  quantum operators with the usual associated indeterminacy and uncertainty. An extended spacetime region emerging from Planck energy waves has a substantial quantum uncertainty, far larger than the Planck length.

 The central hypothesis is that the
 position of any body in  spacetime is defined by some   quantum observable operator, $\hat  x$.  It operates on  quantum states of spacetime that connect measurements made at widely separated events, normally connected by assuming a classical metric.  The following theory describes a  candidate form of the connection between events on null trajectories, by reference to classical null trajectories.

 Consider an operator $\hat x_\perp$ that measures  position along an axis $x$ perpendicular to a classical null trajectory connecting events 1 and 2.  At time $t_1$ a measurement is made of body 1 and at time $t_2$ a measurement is made of body 2.  We can write a quantum-mechanical amplitude for obtaining a particular pair of position results as
 \begin{equation}\label{positionamplitude}
\langle x_{\perp 1}, x_{\perp 2}\rangle
=\langle \psi_{body2}|\hat x_{\perp 2}|\psi_{spacetime}\rangle
\langle \psi_{spacetime}|\hat x_{\perp 1}|\psi_{body1}\rangle.
\end{equation}
The bodies in question may be proof masses in an interferometer, other observer apparatus, black holes, or elementary particles.
The states of the bodies and the spacetime are labled by their wavefunctions $\psi$. In particular, the wavefunctions in the states $|\psi_{body1}\rangle, \langle \psi_{body2}|$ refer to the usual quantum mechanical wavefunctions referring to a  local, classical spacetime position.  This formula represents coherent combinations of spacetimes and positions leading to a given quantum-mechanical amplitude for the specified overall result.  

The analysis below develops a simple model for the effects of the connecting spacetime, $|\psi_{spacetime}\rangle
\langle \psi_{spacetime}|$.  In the case where $\psi_{body 1}$ and $\psi_{body 2}$ are very narrow--- such as macroscopic proof masses in an interferometer---  the structure of $\psi_{spacetime}$   still leads to indeterminacy in position. 
The usual procedure in quantum mechanics and field theory assumes a classical spacetime, so that the state $|\psi_{spacetime}\rangle
\langle \psi_{spacetime}|\hat x_{\perp 1}|\psi_{body1}\rangle$ is represented by a delta-function of deviation from the classical null trajectory from $x_{\perp 1}$.  The amplitude for body 2  to be at a particular position in that case is just given by the wavefunction of the body $ \psi_{body 2}$, and indeed this idea defines the meaning of position wavefunction.  When $\psi_{body2}$ is very narrow, as in a macroscopic body rather than a particle, the position is classical. 
Here we focus on the new effects created by including the width of  $\psi_{spacetime}$.

The measurement process summarized in Eq.(\ref{positionamplitude})
  does not specify a particular form of  interaction, but comprises   an operational definition of position.   The relative position amplitude $\langle x_{\perp 1}, x_{\perp 2}\rangle$    replaces the definite classical positions of bodies or particles in their interactions with other bodies or particles.  The difference is only noticed in nonlocal comparisons of positions.  For those, the final amplitude now depends on the spacetime wavefunction connecting widely separated events,  and   has a width   dominated by the overall envelope of  $\psi_{spacetime}$.  
No new locally detectable interactions or effective fields are predicted on either a classical or quantum level. Instead, there is a loss of precision and determinacy that increases with the scale over which a measurement is made.

The state $|\psi_{spacetime}\rangle$ can be decomposed in the usual way,
\begin{equation}
|\psi_{spacetime}\rangle=\sum a_i|\psi_i\rangle.
\end{equation}
As usual in quantum mechanics, an observation   
  fixes the state to be one of those components at the moment of observation, when it must be in an eigenstate of the observable operator. The measurement of an observable is a process of correlation that fixes a branch of the wavefunction along a future null cone from the observation event.  The arguments here suggest a simple and general physical motivation for a particular kind of decomposition in nearly flat spacetime: the spacetime eigenstates are  a set of approximately plane wavefunctions.  Transverse localization leads to an indeterminacy of orientation for those wavefunctions.

We choose to decompose the spacetime wavefunction using classical null trajectories as reference states, since these define a frame-independent invariant causal structure.
 Consider  a spacetime reference particle chosen  to have Planck energy in a particular reference frame (again, not a tranformation property expected of a regular particle or field).  Its wavefunction will be identified with a particular eigenstate of $\psi_{spacetime}$.   Its classical null trajectory is along the $z$ axis in flat spacetime, with $x=y=0$, and momentum $p_0=\hbar  /l_P$. In the $z$ direction its quantum wavefunction is a  plane wave,
 \begin{equation}
\psi\propto \exp[i(t-z)/l_P]
\end{equation}
However,   specifying a classical trajectory to measure spatial positions (via Eq. \ref{positionamplitude})  requires localization in the transverse directions, $x_\perp$ and $y_\perp$.  As a  result  the orientation of the wave and therefore the classical trajectory are subject to quantum indeterminacy. Then the measured transverse positions of any pairs of bodies as emergently defined in Eq.(\ref{positionamplitude}) will have  wavefunction widths relative to this classical null trajectory determined not mainly by  their locally defined wavefunctions $\psi_{body}$, but that of the reference particle representing the spacetime wavefunction connecting them,
 $\psi_{spacetime}$.

 At  time $t_1$  the reference  particle obeys the usual Heisenberg commutation relation between momentum and position operators along the transverse $x$-axis,
\begin{equation}\label{heisenberg}
[\hat x_\perp(t_1),\hat p_\perp(t_1)]=-i\hbar,
\end{equation}
where   $x_\perp,p_\perp$  refer to any axis chosen perpendicular to the null trajectory connecting events 1 and 2.
Thus the spacetime wavefunction is   a plane wave in $z$ but has an uncertain orientation  associated with  localized measurements in the $x_\perp$ direction.

 Consider  the reference particle at time $t_2$ in a particular frame.  (Recall that this is a null trajectory so the actual spacetime interval between these events vanishes.)
The transverse momentum  $p_\perp(t_1)$ of the  particle at event 1 is related to its transverse position displacement at event 2 by the angular deflection,
\begin{equation}\label{deflection}
 p_\perp(t_1) /p_0= p_\perp(t_1)l_P/\hbar= x_\perp(t_2)/ (t_2-t_1).
\end{equation}
Combining equations (\ref{heisenberg}) and (\ref{deflection}) 
yields a commutation relation between transverse position operators, 
\begin{equation}\label{commute}
[\hat x_\perp(t_1),\hat x_\perp(t_2)]=-i l_P (t_2-t_1).
\end{equation}
This formula specifies in terms of observable operators the effect of the intervening spacetime operator,  $ |\psi_{spacetime}\rangle
\langle \psi_{spacetime}|$,   in Eq.(\ref{positionamplitude}).
Transverse spatial positions  thus  become conjugate variables subject to quantum indeterminacy.

It is interesting that even though Eq.(\ref{commute}) is a quantum commutation relation,  Planck's constant $\hbar$ does not appear explicitly.  Once the conjugate quantum variables are both spatial positions, $\hbar$ is no longer needed, as it is when relating position and momentum operators: the quantum uncertainty is a purely geometrical effect and depends only on the scale of a system compared to the fundamental length $l_P$.    The actual Planck constant  only appears implicitly in connection with other units (and with other fields via $G_N$)  through the usual definition of $l_P$, which according to our  conjecture plays the role of a   minimum length for the fundamental theory in any frame.

It should be emphasized that the transverse momentum here does not refer to an interaction with a particular particle. Rather, it is the transverse momentum associated with the specification of any classical trajectory. Although we have used particle/wave duality to describe a reference particle, this is just a device for estimating the effects of the spacetime wavefunction, corresponding to departures from a definite classical null path. The plane wave is coherent in the transverse direction: nearby paths share almost the same value of deviation $x_\perp$ from a local reference trajectory. Thus the indeterminacy cannot be detected locally even in the transverse direction. Clearly this transverse coherence of an effective spacetime wavefunction cannot   extend indefinitely but for consistency must decohere slowly with transverse separation, as discussed below.
\section{holographic uncertainty of position and angle}
In the usual way, the indeterminacy (Eq. \ref{commute}) yields a Heisenberg uncertainty relation,
\begin{equation}\label{uncertainty}
\Delta x_\perp(t_1)\Delta x_\perp(t_2)>l_P (t_2-t_1)/2,
\end{equation}
where $ \Delta x_\perp(t_1),\Delta x_\perp(t_2)$ denote the standard deviations at events 1 and 2 of the distributions of position measurements of any body or particle, describing quantum-gravitational departures from the classical null ray connecting them.  The standard deviation $\Delta x_\perp$ of the  difference in relative transverse positions   is then given by $\Delta x_\perp^2=  \Delta x_\perp(t_1)^2+\Delta x_\perp(t_2)^2$; it has a minimum value when $ \Delta x_\perp(t_1)=\Delta x_\perp(t_2)$,
so 
\begin{equation}\label{uncertainty2}
\Delta x_\perp^2>l_PL:
\end{equation}
a ``holographic uncertainty principle'' for   transverse positions   at events of null spacetime separation and spatial separation $L$ in the measurement frame.  
Since the wave direction $z$ defines any classical null trajectory, we interpret this as uncertainty associated with $\psi_{spacetime}$,   a  quantum indeterminacy of the metric of a  spacetime emerging from quantum behavior of Planck-scale quantum waves. 

From this we can also derive a minimum uncertainty in angular orientation of any null ray of length $L$.  The angular departure from the  classical reference null ray is $x_\perp(t_2)- x_\perp(t_1)$, so the standard deviation $\Delta\theta_x$  of the distribution of angular orientations in the $x$ direction is given by
\begin{equation}
\Delta\theta_x^2= {(\Delta x_{\perp 1}^2 + \Delta x_{\perp 2}^2)\over(t_2-t_1)^2}.
\end{equation}
 This has a minimum value when $\Delta x_\perp(t_1)=\Delta x_\perp(t_2)$, and 
  \begin{equation}
\Delta x_\perp(t_1)^2=\Delta x_\perp(t_2)^2>{l_P (t_2-t_1)/2},
\end{equation}
so we have a lower bound for standard deviation  in the distribution of angles, an uncertainty principle for the angular orientation of any null ray that applies independently to  each transverse axis:
\begin{equation}\label{angledelta}
\Delta \theta_x> \sqrt{l_P/L},\ \  \Delta \theta_y>  \sqrt{l_P/L}.
\end{equation}
 Note that the angular uncertainty actually increases with smaller $L$.  Indeed, one view of emergent spacetime suggested by holographic indeterminacy is that a classical spatial direction is actually ill defined at the Planck scale and only becomes well defined after many Planck lengths of propagation.

  The quantum deflections of neighboring null trajectories are not drawn from independent distributions.  Although the orientation is indeterminate, the same orientation of the  plane wave from event $1$, and the same transverse displacement,  are shared coherently  by all particles on nearby null trajectories. Local measurements of transverse positions are not limited in precision; they all share the same spacetime eigenstate, with local classical spacetime behavior,   as required by agreement with local physics.  In Eq.(\ref{positionamplitude}), the operator $|\psi_{spacetime}\rangle
\langle \psi_{spacetime}|$    operates coherently on measurements of any and  all bodies near the same events.

Spacetime states of  nearby  null trajectories do however  gradually decohere with larger separation. 
Consider two nearby null rays from event 1, terminating ``at the same time''   at distance $L$, with transverse positions  separated by a classical distance $\delta x_{\perp}$.  This   relative transverse position
is indeterminate and uncertain with standard deviation, 
\begin{equation} \label{coherence}
\Delta (\delta x_{\perp})> 
\left({\delta x_{\perp}\over L}\right) \sqrt{l_P\delta x_{\perp}}=
\left({\delta x_{\perp}\over L}\right)^{3/2}\sqrt{l_PL}.
\end{equation}
One way to see this is to consider the effect of spacetime indeterminacy on classical spacelike surfaces of simultaneity,  which also are defined by the emergent metric.  A ray connecting bodies at the classical separation $\delta x_{\perp}$ has its own transverse uncertainty $\simeq \sqrt{l_P \delta x_{\perp}}$, with  a corresponding indeterminacy in the relative distances along the direction to event 1.  For consistency the emergent metric must have a corresponding small uncertainty in the orientation of the constant-time surface,  and therefore in the difference $\delta x_{\perp}$ between positions  assigned to those events.  

Similarly,  consider two nearby ``parallel'' null rays with transverse separation $\delta x_\perp(t_1)$;  the uncertainty in their relative angular orientation after propagating distance $L$ is
\begin{equation}
\Delta\theta(\delta x_\perp)>\left({\delta x_\perp\over L}\right)^{3/2}\sqrt{l_P/L}.
\end{equation}
Holographic uncertainty adds quantum noise to the parallel postulate of Euclidean space; at this level it is indeterminate whether or not rays are parallel.  
 
 The decoherence   leads to detectable quantum position noise, as discussed below. It  can be measured by an interferometer that  compares the relative transverse positions of two widely separated bodies, at a time corresponding to a null separation of measurement events, from another body at a large transverse distance.
    
\section{Spacetime Quantum Degrees of Freedom}
The uncertainty in   orientations or angles translates into a limit on the number of distinguishable angular orientations of plane-wave modes of any quantum field, and therefore a limit on the total  number of quantum degrees of freedom more severe than that  in standard field theory with a cutoff at the Planck scale. 
Taking both transverse directions into account, and counting states by assuming Nyquist sampling on a sphere (that is, states or degrees of freedom separated by two standard deviations in each transverse direction $x$ and $y$), the number of distinguishable orientations for rays   of length $L$ is given by 
\begin{equation}
N={{4\pi L^2}\over{2(\Delta x_{\perp 1}^2 + \Delta x_{\perp 2}^2)^{1/2}\times 2(\Delta y_{\perp 1}^2 + \Delta y_{\perp 2}^2)^{1/2}}}. 
\end{equation}
As above, from Eq. (\ref{uncertainty}), in both $x$ and $y$ directions the uncertainty is minimized for 
$\Delta x_{\perp 1}=\Delta x_{\perp 2}=\sqrt{l_PL/2}$;
the maximum number of distinguishable directions in a sphere of  radius $L$ is thus 
\begin{equation}
N_\theta(L)<{{4\pi}\over{4\Delta \theta_x \Delta \theta_y}}=\pi L/l_P.
\end{equation}
We can take this number as a bound on   the number of distinguishable wavevector directions for field modes confined  to the volume. In field theory, the number of distinguishable wavevector directions in the same volume is  much larger, $\approx 4\pi (L/l_P)^2$. The different behavior shows the effect of  holographic blurring at macroscopic distances: in emergent spacetime, extended field states lack independence they have in field theory.

The number of quantum degrees of freedom up to the Planck scale is  $\approx L/l_P$ field modes per direction, so the number of degrees of freedom is bounded by
\begin{equation}
S\approx N_\theta(L)L/l_P<\pi (L/l_P)^2.
\end{equation}
This   maximum number of degrees of freedom agrees (up to a numerical factor of the order of unity, depending on the exact nature of the Planck cutoff) with the maximum number of degrees of freedom allowed by covariant entropy bounds \cite{Bousso:2002ju}.  Although the derivation has been different, the physical connection is clear:  for example,  the transverse envelopes of the spatial wavefunctions of these spacetime states  resemble those corresponding to particles evaporating from a black hole of Schwarzschild radius $\approx \Delta x_\perp (L)$ into flat space, one process used to estimate covariant entropy bounds. Those particles must experience a similar ``blurring'' in their transverse positions far away, otherwise   their states would contain more information (and observables could reveal more data) than available in the black hole that produces them, a violation of quantum unitary evolution\cite{Hogan:2007hc}.

Although this picture seems radical,   it is also a rather straightforward expression, in terms of behavior in 3-space, of the effects of  holographic  bounds on  spacetime degrees of freedom.     The conjectured holographic behavior of quantum gravity  has long been  supported by arguments based on  black hole physics and string theory\cite{'tHooft:1985re,Strominger:1996sh,'tHooft:1993gx,Susskind:1994vu,'tHooft:1999bw,Bigatti:1999dp,Bousso:2002ju,Maldacena:1997re,Witten:1998zw,Aharony:1999ti,Alishahiha:2005dj,Horowitz:2006ct,Padmanabhan:2006fn}, in spite of its apparent contradiction with features of  field theory such as locality. 
 The detailed quantitative agreement is the main reason to adopt $l_P$  as the fundamental length for this theory, and suggests that the decomposition into transversely-localized plane waves approximates a complete decomposition of the quantum degrees of freedom of nearly-flat spacetime.

 \section{observable holographic shear noise}
 
 This implementation of holography has  physical implications for observable real-world  behavior.  Transverse positions  at separation  $\simeq L$ in any frame, measured from a distance $\simeq L$,  have an  uncertainty   
   $\Delta x_\perp> \sqrt{l_PL }$, which is $> \sqrt{L/l_P}$ times larger than the local   resolution of Planck scale field theory.  Similarly, the relative orientation of  null geodesics of length $L$ and separated by $L$  has an irreducible angular uncertainty  $\Delta \theta > \sqrt{l_P/L}$ in each orthogonal direction.  For macroscopic $L$, these uncertainties brings Planck scale quantum gravity within reach of realistic direct experiments on positions of macroscopic masses.

Holographic  indeterminacy predicts a new kind of observable quantum noise that may be thought of  as a kind of sampling noise that leads to an apparent Brownian motion of spacetime.   As we have seen, measurement of relative transverse position
of two objects separated by macroscopic distance $L$, at events separated by a null trajectory, yields an indeterminate result. This property implies that measurements of relative transverse positions show a new source of random noise  that increases with spatial and temporal separation like $\sqrt{l_PL}$. 

This phenomenon mimics  a new source of gaussian  random noise in spatial shear modes of metric perturbations.  It can be  detected by measurements of classical positions where it appears as a new source of quantum noise, caused not by the usual quantum noise associated with particles and fields, but with quantization of the emergent  spacetime metric.  Because of the holographic scaling it is called  ``holographic noise.'' 

The noise appears even in some combinations of purely radial distance measurements.  To see this, consider bodies at rest arranged in a right triangle with sides measured by interferometry.  Classically, the  complete shape of the triangle is determined by measurements between events separated by null trajectories.  On the other hand the orthogonal transverse positions of the bodies adjacent to the hypotenuse are indeterminate, which affects the measured length of the hypotenuse.  The shape of the triangle is therefore also indeterminate.  Once it is measured, it is placed in an eigenstate and  does not immediately completely decohere, so subsequent measurements yield a similar result.  However, after a decoherence time, roughly a light travel time around the triangle,   the relative positions are again indeterminate. The measured radial distances between  the bodies change accordingly since they inhabit a common branch of the wavefunction corresponding to a  classical spacetime.   

The transverse spatial character of holographic shear noise distinguishes it from gravitational wave strain, even though we can use the same technology to detect both types of perturbation and the same units to describe both. 
 The metric perturbation of a   gravitational wave with $+$ or $\times$ polarization, propagating on the $z$ axis, has a 3-space dependence in the transverse-traceless (TT) gauge,
\begin{eqnarray}
 h^+_{ij} =
h^+ \left(
\begin{array}{ccc}\label{strain}
1  & 0  &  0 \\
 0 & -1  &  0 \\
  0 &  0 &    0
\end{array}
\right),\qquad
 h^\times_{ij} =
h^\times \left(
\begin{array}{ccc}
0  & 1  &  0 \\
 1 & 0  &  0 \\
  0 &  0 &    0
\end{array}
\right).
\end{eqnarray}
(Here the indices $i,j=1,2,3$ correspond to $x,y,z$. The full description of the wave multiplies these by  the usual wave dependence, $e^{-i(kz-\omega t)}$.)
By contrast the apparent metric perturbations of   $x$- and $y$-polarized   shear modes propagating on the $z$ axis have a 3-space dependence,
\begin{eqnarray}
h^x_{ij} =
h^x_H \left(
\begin{array}{ccc}\label{shear}
0  & 0  &  1 \\
 0 & 0  &  0 \\
 -1 &  0 &    0
\end{array}
\right),
\qquad
h^y_{ij} =
h^y_H \left(
\begin{array}{ccc}
0  & 0  &  0 \\
 0 & 0  &  1 \\
 0 &  -1 &    0
\end{array}
\right).
\end{eqnarray}
The components of spatial displacement are
\begin{equation}
\Delta x^j={1\over 2} h_{jk}x^k,
\end{equation}
so as expected the shear perturbation corresponds to the displacement caused  by a change in the orientation of $z$. Classically there is no physical significance to shear perturbations of flat space since they can be removed by coordinate transformation.  However, holographic indeterminacy adds a quantum stochastic element to transverse positions, so even flat space has an apparent gaussian superposition of these perturbations with random phases and orientations.   It can be characterized by the power spectral density of the metric perturbation amplitude $h_H$ as function of frequency.

The spectrum of the holographic noise is universal and depends only on the one scale in the system,  $l_P$.  The power spectral density (mean square dimensionless metric perturbation per frequency interval)  is independent of frequency,  given by\cite{Hogan:2007hc}
\begin{equation}
S_H\simeq l_P/c.
\end{equation}
In   units similar to those used to describe equivalent metric strain noise in gravitational wave detectors, the rms  amplitude of holographic noise in terms of equivalent metric shear is
\begin{equation} \label{noise}
h_{H,rms}=S_H^{1/2}\simeq  \sqrt{l_P/c} = 2.3 \times 10^{-22} /\sqrt{\rm Hz}.
\end{equation}
   This   noise level is comparable to that already achieved for measurements of  tensor waves in  science runs of operational interferometers such as LIGO\cite{Abbott:2006zx}.  Holographic noise should therefore be  measurable with similar technology.  However, it can only be detected in a system designed with a layout capable of   detecting transverse positions at events with  macroscopic spatial separation.

  Like gravitational waves, these perturbations cannot be detected in a purely local experiment. 
However, unlike the tensor modes of physical gravitational waves (Eq.\ref{strain}),   holographic perturbations in position are transverse to  spatial separation (Eq. \ref{shear}).  They can never be detected, as gravitational waves can, by   purely radial measurements in one direction.  (Otherwise it would be possible to extract energy from  the holographic noise, which is impossible.) They can however be detected by comparing the results of measurements of different paths.  The experiment must not only be nonlocal but must also compare transverse positions  at events with null separation, widely separated in space (in the experiment frame).  For example, a Michelson interferometer like LIGO compares transverse distances in two arms, but only at one place,   so it does not detect holographic noise.
The indeterminacy produced by the holographic shear noise can however be detected in  gravitational-wave detector interferometers of appropriate geometry.  Purely radial measurements of distances along some baselines can  be arranged to measure transverse positions associated with other baselines at wide separations. Some of the experimental options for achieving this are discussed in \cite{Hogan:2007hc}.  They include triangular configurations, such as LISA.  Holographic noise for example leads to   random variations in the total pathlength in a circuit that can be detected in a Sagnac-type interferometer, even if it is synthesized from three separate radial distance measurements\cite{Hogan:2007hc,armstrong,Hogan:2001jn,Cornish:2001bb,sylvestre,Tinto:2004yw}.  The coherence estimate above (Eq. \ref{coherence}) suggests that the detectable noise is limited by the shortest side of the  triangle. 

Holographic indeterminacy appears to be a new feature of emergent quantum spacetime with no physical  counterpart in classical gravity.   In perturbations of classical  flat spacetime these shear modes are simply gauge artifacts; they carry neither energy nor information, and have no observable effects.  This feature partially explains why quantized shear modes have not previously  played a more prominent role in quantized gravity in field theory.  Holographic noise indeed does not carry energy or information, but it does appear to create an observable nonlocal effect associated with quantum indeterminacy in the branching of the spacetime metric.

 Strange as holographic noise seems, it is hard to reject out of hand the idea that there must be some nonlocal indeterminacy in position, without  some equally profound modification of quantum mechanics (e.g., \cite{'t Hooft:1999gk}).  The noise is a direct consequence of information limits and in that sense is more general than the particular indeterminacy model described here.  The  holographic entropy bounds (if valid)  require that there must be fundamental limits on the dimension of Hilbert space and therefore a limit on the overall precision of position measurements, creating an effective granularity much coarser than the Planck resolution implied by field theory.  These must be manifested in some form of  added effective noise to measurements by an apparatus capable of measuring positions to sufficient accuracy in a two-dimensional plane. Consider again a right triangle:  without holographic indeterminacy, it would be possible to measure the   positions of the vertex events to Planck precision,  but holography does not allow enough information in the spacetime to specify such a  large number of distinguishable positions with all the independence allowed by field theory. 
 The limited number of independently distinguishable positions or angular orientations necessarily translates into a sampling noise in observables.   If holographic bounds are  implemented by holographic indeterminacy, they  lead inevitably  to  holographic noise\cite{Hogan:2007hc}; other implementations of holography may produce less conspicuous effects, but they necessarily imply the same information deficit.
 
  Although no new particles are predicted from this theory, indeterminacy  might measurably affect the behavior of  some systems even on a microscopic scale.  Nonlocality, relative to Planck distances, applies  even on a subatomic scale.  Spatially extended, coherent quantum states sample a spacetime that has holographic quantum variations in its geometry.  For example, the background classical spacetime for an extended  state such as an atomic nucleus, with a typical  length scale $\approx 10^{20}l_P$, has angular uncertainty along each axis of   $\Delta \theta\approx 10^{-10}$; for atomic or molecular states, $\Delta \theta\approx 10^{-13}$. Although   symmetries  in some  atomic and nuclear  coherent eigenstates are tested more precisely than this, spatially extended coherent eigenstates that  would  be coherent in classical spacetime should remain so in the presence of holographic spacetime indeterminacy; the intermediate states $|\psi_{spacetime}\rangle
\langle \psi_{spacetime}|$  in Eq. (\ref{positionamplitude}) do not introduce noise if the whole system is in a coherent eigenstate since no measurement is made collapsing the spacetime to a definite eigenstate of transverse position.    On  the other hand, fields interact on a  fluctuating geometry, which might allow coherent mixing or other observable effects in systems where the   Hamilitonian  has some dependence on transverse position.  It is worth considering whether some microscopic quantum systems  could be found or constructed with  a configuration   that has measurable consequences of holographic indeterminacy.  

\bigskip
 \acknowledgements
The author is grateful for support from  the Alexander von Humboldt Foundation and the hospitality of the Max-Planck-Institut f\"ur Astrophysik.

\end{document}